\documentclass[aps,prb,reprint,floatfix,amsmath,amssymb,graphicx,longbibliography,showpacs]{revtex4-1}

\usepackage{amssymb,amsmath}
\usepackage{graphicx}
\usepackage{subfigure}
\usepackage{bm}
\usepackage[varg]{txfonts}
\usepackage{color}

\def\be{\begin{equation}}
\def\ee{\end{equation}}
\def\bea{\begin{eqnarray}}
\def\eea{\end{eqnarray}}
\def\bmatrix{\begin{pmatrix}}
\def\ematrix{\end{pmatrix}}

\newcommand{\bk}{ \mathbf{k} }

\newcommand{\bq}{ \mathbf{q} }
\newcommand{\bqq}{ \mathbf{Q} }

\begin{document}

\title{Hall and Seebeck coefficients from bi-directional charge density wave state in high-$T_c$ cuprates }

\author{Kangjun Seo}
\affiliation{School of Natural Sciences, University of California Merced, Merced, California 95343, USA}
\author{Sumanta Tewari}
\affiliation{Department of Physics and Astronomy, Clemson University, Clemson, SC 29634, USA}

\date{\today{}}

\begin{abstract}
The recent discovery of an incipient charge density wave (CDW) instability competing with superconductivity in a class of high temperature cuprate superconductors has brought the role of charge order in the cuprate phase diagram under renewed focus. Here we take a mean field $Q_1=(2\pi/3,0)$ and $Q_2=(0,2\pi/3)$ bi-axial CDW state and calculate the Fermi surface topology and the resulting Hall and Seebeck coefficients as a function of temperature and doping. We establish that, in the appropriate doping ranges where the low temperature state (in the absence of superconductivity) is a bi-directional CDW, the Fermi surface consists of electron pockets, resulting in the Hall and Seebeck coefficients becoming negative at low temperatures as seen in experiments.
\end{abstract}

\pacs{74.25.F-,71.45.Lr,74.72.-h}

\maketitle


\section{Introduction}
Despite intensive efforts in the last two decades understanding the nature of the pseudogap phase in cuprate high temperature
superconductors remains an open problem \cite{Norman:2005}. At zero hole doping, the insulating parent compounds of these systems are three-dimensional (3D) antiferromagnetic insulators.
As holes are introduced, the superconducting transition temperature ($T_c$) is finite above a critical doping, reaching a maximum at a hole concentration  known as optimal doping. The regime with hole doping less than optimal -- known as underdoped regime -- hosts the remarkable pseudogap phase. In this regime, the normal (non-superconducting) phase above $T_c$ has an anisotropic spectral gap of unknown origin (`pseudo'-gap), and behaves strikingly differently from a Fermi liquid. It is generally understood that the crux of the problem of $d$-wave pairing resides in the pseudogap phase, from which superconductivity develops at lower temperatures.

The idea that in the underdoped regime of the cuprates various charge, spin, or current ordered states compete with superconductivity has been apparent in the past few years~\cite{Varma:1997,Chakravarty:2001,Kivelson:2003,Hosur:2012,Orenstein:2012,Chakravarty:2013,Wu:2011,Wu:2013}.
A compelling evidence is provided in the class of materials YBa$_2$Cu$_3$O$_{6+x}$ (YBCO), in which, upon suppression of superconductivity by a strong magnetic field, exquisite quantum oscillations of various electronic properties with the applied field point to the existence of Fermi surface pockets \cite{Doiron-Leyraud:2007,Sebastian:2008}. Since in the overdoped regime the Fermi surface is large and hole-like, the existence of Fermi pockets in the underdoped regime indicates a Fermi surface reconstruction near optical doping. A change in the topology of the Fermi surface from hole-like at higher doping to electron-like at low doping is also indicated by measurements of low temperature Hall and Seebeck coefficients which turn negative in the underdoped regime \cite{LeBoeuf:2007,Chang:2010,Laliberte:2011}.
Although various charge, spin, and current ordered states have been proposed to account for the Fermi surface reconstruction in YBCO \cite{Chakravarty-Kee,Millis-Norman}, none had so far been observed in bulk-sensitive probes until recently.

In recent X-ray diffraction experiments, two groups have independently discovered \cite{Chang:2012,Ghiringhelli:2012,Kivelson,Tranquada} strong evidence for an
incipient charge density wave (CDW) instability (with scattering peaks at wave vectors $Q_1=(q_1,0,0.5)$ and $Q_2=(0,q_2,0.5)$, with $q_1 \sim q_2 \sim 0.31$) competing with superconductivity in a range of doping in the underdoped regime of YBCO. Although not conclusively known if the peaks derive from equal distributions of domains with only uni-axial CDW correlations (i.e., with modulations given by only $Q_1=(q_1,0,0.5)$ or $Q_2=(0,q_2,0.5)$)  or a bi-axial CDW with both wave vectors co-existing, the lack of anisotropy in the scattering signals (intensities, widths) indicates a coupling between $Q_1$ and $Q_2$ and a bi-axial CDW order. The temperature ($T$) dependent correlation length above $T_c$ and only short ranged correlations perpendicular to the CuO$_2$ planes indicate that the observed charge order is only quasi-static. However, the near divergence of the correlation length as $T \rightarrow T_c$ (and that the scattering signals significantly increase on application of magnetic field below $T_c$) indicate that a true thermodynamic CDW transition at some critical temperature ($T_{CDW}<T_c$) is preempted by the superconducting transition at $T_c$. Evidence for a similar charge density wave transition in the underdoped regime has also been found in other recent experiments~\cite{LeBoeuf:2013,Hinton:2013,sebastian:2014,Fujita:2014,Comin:2014}.

In this paper we treat a two-dimensional (2D) bi-axial $Q_1=(2\pi/3,0)$ and $Q_2=(0,2\pi/3)$ CDW state in mean field theory (valid for temperatures $T<T_{CDW}$ and magnetic fields high enough to eliminate the superconductivity) and investigate the reconstructed Fermi surface and experimentally measured transport coefficients as functions of temperature and hole doping appropriate for the underdoped regime of the cuprates. Although the experimental evidence is that for a slight incommensuration in the CDW scattering vectors (i.e., $q_1 \sim q_2 \sim 0.31$) in this paper we work with a commensurate CDW for simplicity (i.e. we take $q_1=q_2=0.33$, corresponding to charge modulations with periodicity of three lattice vectors). We find that, below the CDW transition temperature and in the appropriate regime of hole doping, the Fermi surface topology changes from a large hole-like Fermi surface at higher doping to a few
electron and hole-like Fermi pockets in the CDW state.
A similar fermi surface reconstruction in terms of a different CDW state was recently assumed to explain the low frequency of quantum oscillations in the pseudogap phase of the cuprates~\cite{sebastian}.
An explanation of the Fermi pockets observed in the recent quantum oscillation experiment in Ref.~\cite{sebastian:2014} in terms of the bi-axial CDW state considered in the present work is left for future study.
We calculate the CDW Hall and Seebeck coefficients using the Kubo formula which reduces to the Boltzmann expressions in the experimentally appropriate limits. We show that, in the appropriate doping range where the low temperature phase (in the absence of superconductivity) is a bi-axial CDW, the Hall and Seebeck coefficients turn negative for $T<T_{CDW}$, as seen in the experiments. Our calculations establish that the recent observations of strong incipient CDW instability in YBCO \cite{Chang:2012,Ghiringhelli:2012} are broadly consistent with the other recent experimental surprises in underdoped regime, namely, the existence of Fermi surface pockets \cite{Doiron-Leyraud:2007,Sebastian:2008} and negative low temperature transport coefficients indicating Fermi surface reconstruction \cite{LeBoeuf:2007,Chang:2010,Laliberte:2011}.

The paper is organized as follows. In Sec.~\ref{secii}, we calculate the Fermi surface in the presence of a bi-axial $Q_1$ and $Q_2$ CDW order parameter in mean-field theory. We present in Sec.~\ref{seciii} the temperature and doping dependent crossover in sign of the Hall and Seebeck coefficients using the Kubo formula. In Sec.~\ref{seciv}, we show that such sign changes in the coefficients does not occur in a uni-axial CDW state. We summarize and conclude in Sec.~\ref{secv}.

\begin{figure}[t]
\includegraphics[width = 1\linewidth]{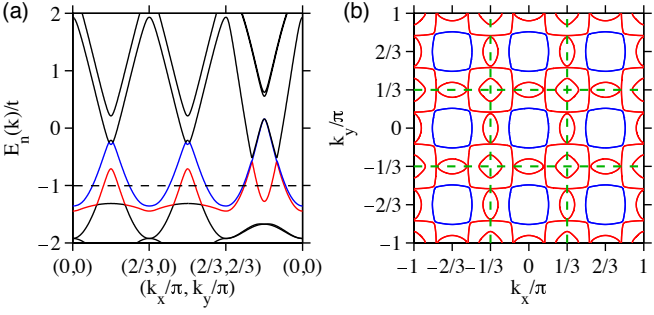}
\caption{
(color online)
{(a) The CDW band structure from the eigenvalues $E_n(\bk)$ of the matrix $H(\bk)$, Eq.~(\ref{h_matrix}). The chemical potential, corresponding to hole doping $p=0.125$, is noted by a dashed line. Here, the CDW order parameter $W = 0.069$ eV at $T=0$.
(b) The Fermi surface for $p=0.125$ in the first Brillouin zone. The green dashed square in the middle represents the reduced Brillouin zone (RBZ) boundary determined by the CDW momenta $Q_1$ and $Q_2$. Note that the Fermi surface in the CDW state consists of two electron pockets in RBZ: one large electron pockets at the center $(0,0)$ and the corners $(\pi/3,\pi/3)$ of the RBZ (red) and two hole pockets at the boundary $(\pi/3,0)$ and $(0,\pi/3)$ (red).
}
}
\label{fermi_surface_cdw}
\end{figure}

\section{Mean-field Hamiltonian and CDW Fermi surface\label{secii}}
To calculate the Fermi surface and the normal-state quasiparticle Hall and Seebeck coefficients, we consider electrons on a two-dimensional square lattice of unit lattice constant, with a dispersion given by
\bea
\label{dispersion}
\varepsilon_\bk
&=&
-2t \left( \cos k_x + \cos k_y  \right) +
4 t^\prime \cos k_x  \cos k_y
\nonumber\\
&&-2t^{\prime\prime} \left( \cos 2 k_x + \cos 2 k_y \right),
\eea
where $t$ is a nearest-neighbor, $t^\prime$ is a next-nearest-neighbor, and $t^{\prime\prime}$ is a next-to-next-nearest-neighbor hopping integrals. For the Fermi surface of YBCO measured in photoemission experiments, we choose the parameters $t=0.3$ eV, $t^\prime = 0.32 t$, and $t^{\prime\prime} = 0.1 t^\prime$.
Within a mean-field picture, the Hamiltonian for the CDW state is described by the quasiparticles subject to a periodic modulation with wave vector $Q_1=(2\pi/3,0)$ and $Q_2=(0,2\pi/3)$,
\be
H_{MF}
=\sum_{\bk,\sigma} \left( \varepsilon_\bk - \mu \right) c_{\bk\sigma}^\dagger c_{\bk\sigma}+
\sum_{\bk,Q_i,\sigma} W_{Q_i} c_{\bk+Q,\sigma}^\dagger c_{\bk\sigma} + H.c.,
\ee
where $c_{\bk\sigma}$ is the annihilation operator for an electron of spin $\sigma$ and momentum $\bk$, and
$
W_{Q_i} = U\sum_{\bk} \langle c_{\bk+Q_i, \sigma}^\dagger c_{\bk, \sigma} \rangle
$ is a CDW order parameter with on-site Coulomb repulsion $U$.
By taking $W_{Q_1} = W_{Q_2} =  W$, we can express the Hamiltonian in terms of a nine-component quasiparticle spinor $\psi_\sigma(\bk)$ as
\be
H_{MF} =
\sum_{\bk \in \text{RBZ},\sigma}
\psi_\sigma^\dagger(\bk) H(\bk) \psi_\sigma(\bk),
\ee
where the reduced Brillouin zone (RBZ) is given by $|k_x|, |k_y|\le \pi / 3$, and  $H(\bk)$ is $9\times 9$ matrix
\begin{widetext}
\be
\label{h_matrix}
H(\bk) =
\bmatrix
\xi_{\bk} & W & W & W & 0 & 0 & W & 0 & 0\\
W & \xi_{\bk+Q_1} & W & 0 & W & 0 & 0 & W & 0\\
W & W & \xi_{\bk-Q_1} & 0 & 0 & W & 0 & 0 & W\\
W & 0 & 0 & \xi_{\bk+Q_2} & W & W & W & 0 & 0\\
0 & W & 0 & W & \xi_{\bk+Q_1+Q_2} & W & 0 & W & 0 \\
0 & 0 & W & W & W & \xi_{\bk-Q_1+Q_2} & 0 & 0 & W \\
W & 0 & 0 & W & 0 & 0 & \xi_{\bk-Q_2} & W & W \\
0 & W & 0 & 0 & W & 0 & W & \xi_{\bk+Q_1-Q_2} & W \\
0 & 0 & W & 0 & 0 & W & W & W & \xi_{\bk-Q_1-Q_2}
\ematrix.
\ee
\end{widetext}
The energy spectrum $E_n(\bk)$ and corresponding eigenoperators $\chi_n(\bk)$ can be obtained by diagonalizing the matrix $H(\bk)$.
Note that the chemical potential has to be adjusted for the value of $W$ to preserve the doping $p$.
The number of electrons per unit cell equals $2N$ due to spin degeneracy, and the doping of the cuprates is counted from half-filling, hence $p = 1 - 2N\,$, where $N$ is the occupied part of the Brillouin zone
\be
N = \sum_{\bk \in \text{RBZ},n} f(E_n(\bk))\,,
\ee
where $f(x) = 1/(1+\exp (x/T)$ is a Fermi distribution function. We find that at $T=0$, $\mu = -0.6989 t$ corresponds to half-filled ($p=0$) and $\mu = -1.0027 t$ corresponds to $p = 0.125$. In Fig.~\ref{fermi_surface_cdw}, we plot the band structure and the Fermi surface corresponding to the CDW state from the eigenvalues $E_n(\bk)$ for hole doping $p=0.125$ in the first Brillouin zone. We find that the Fermi surface consists of electron pockets as well as hole pockets in the reduced Brillouin zone given by the green dashed line. Note that there are two electron pockets in RBZ: one electron pocket at the center $(0,0)$ and the other at the corners $(\pi/3,\pi/3)$ of the RBZ (blue and red), and two hole pockets at the boundary $(\pi/3,0)$ and $(0,\pi/3)$ (red).  Since for higher doping, the free electron Fermi surface in the absence of CDW (given by $\varepsilon_{\bf k}$ in Eq.~(\ref{dispersion}) set to the chemical potential) is large and hole-like, the electron pockets in Fig.~\ref{fermi_surface_cdw} at lower doping indicate a
doping-induced Fermi surface reconstruction.

\section{Hall and Seebeck Coefficients\label{seciii}}
In linear-response theory, three conductivity tensors $\hat \sigma$, $\hat \alpha$, and $\hat \kappa$ are related with charge current $\vec J$ and thermal current $\vec Q$ such as
\be
\bmatrix
\vec J \\ \vec Q
\ematrix
=
\bmatrix
\hat \sigma & \hat \alpha \\
T \hat \alpha & \hat \kappa
\ematrix
\bmatrix
\vec E \\ - \vec \nabla T
\ematrix.
\ee
The electric field induced by a thermal gradient in the absence of an electric current $\vec J = 0$ is given by $\vec E = \hat \sigma^{-1} \hat \alpha \vec \nabla T$. Therefore, the Seebeck coefficient $S$, defined as the $E_x$ generated by a thermal gradient $\nabla_x T$, reads as, $S = \alpha_{xx}/\sigma_{xx}$.
The Hall coefficient, defined as the transverse $E_{y}$ generated by a magnetic field $B$, is given by $R_H = \sigma_{xy}/\sigma_{xx}^2$.

\begin{figure}[t]
\includegraphics[width = \linewidth]{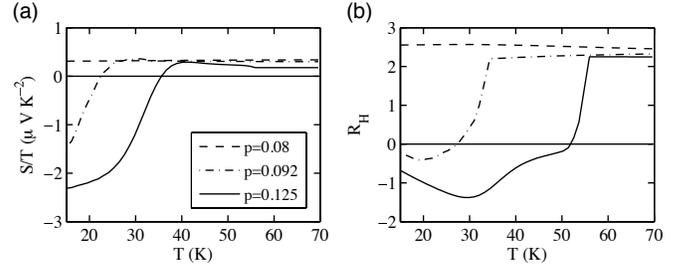}
\caption{
Temperature dependence of the Seebeck and Hall coefficients in the CDW state.
(a) Normal-state Seebeck coefficients $S/T$ at various doping concentrations $p$ are plotted as a function of temperature $T$.
The negative Seebeck coefficients at low temperature for $p=0.092$ and $0.125$ are ascribed to the electron pockets in the Fermi surface due to the CDW, while $S/T$ at $p=0.08$ ($W=0$) remains positive.
(b) Hall coefficients $R_H$ in arbitrary units for various $p$ are plotted as a function of $T$.
The sign of the Hall coefficient changes at lower temperature for $p=0.092$ and 0.125, while it remains positive at all temperatures for $p=0.08$.
}
\label{fig-temp_dep}
\end{figure}

We calculate the dc transport properties using the Kubo formula by applying perturbative electric and magnetic fields: $\vec E = E_0 \hat x \cos \omega t$, $\vec B = q \hat z A_0 \sin q y$.
Further, we assume that the scattering time $\tau(E,\bk)$ is independent of the energy, but momentum dependent: $\tau(E,\bk) = \tau(\bk)$.
Phenomenologically, we retain a possible momentum dependence of the scattering time because, as has been shown before, an assumption of a momentum independent $\tau$ results in a negative Seebeck coefficient in the normal state~\cite{hildebrand, kontani, storey} (i.e., above the CDW transition temperature), inconsistent with experiments. In this work we assume the scattering time $\tau(\bk) = ( 1 + \alpha  ( \cos k_x + \cos k_y ) )^2$ with $\alpha=0.4$. This form of $\tau(k)$ is purely phenomenological and designed to enhance the contribution of the free fermion hole pockets at high temperatures (i.e., above the CDW transition temperature) located near $(\pi,\pi)$ and symmetry related points in the first Brillouin zone. By enhancing the contribution of the hole pockets the high temperature ($W=0$) Seebeck coefficient is positive, in accordance with the experiments~\cite{Chang:2012,Ghiringhelli:2012}. The precise functional
form of $\tau(k)$ is unimportant, however, and any other momentum dependence of the scattering time that enhances the contribution of the high temperature hole pockets to the Seebeck coefficient works just as well. Let us also emphasize that the drop in the Seebeck and Hall coefficients with decreasing $T$ and the eventual negative sign at low temperatures (Fig.~\ref{fig-temp_dep}) are both robust features of the bi-axial CDW state and are immune to variations in the precise functional forms of the scattering time. In particular they continue to hold even with a $k$-independent $\tau$.

To lowest order in the applied fields $\vec{E},\vec{B}$,
the conductivity $\sigma_{xx}$ is given by the Fourier transform of the imaginary time-ordered current-current correlation function, $$\sigma_{xx} = \lim_{q\to 0} \frac{1}{\omega} \text{Im}\int^{1/T} d\tau e^{i(\omega+i\delta)\tau} \langle T_\tau J_x(q,\tau)J_x(q,\tau)\rangle,$$ and
$\sigma_{xy}$ is given by,
\bea
\sigma_{xy} &=& \lim_{q\to 0}\frac{B}{\omega q} \text{Re} \int_0^{1/T} d\tau d\tau^\prime e^{i(\omega+i\delta)\tau} \nonumber\\\nonumber
&\times &
\langle
T_\tau J_y(q,\tau) J_x(0,0) J_y(-q,\tau^\prime)
\rangle,
\eea
where $J_i = e \sum_{\bk,n} \chi_n^\dagger(\bk) \frac{\partial E_n(\bk)}{\partial k_i} \chi_n(\bk)$ is the electric current operator.
The thermal conductivity coefficient $\alpha_{xx}$ is given by the appropriate correlation function of the thermal current $\vec Q$ and the electric current $\vec{J}$, $$\alpha_{xx} = \lim_{q\to 0} \frac{1}{\omega} \text{Im}\int^{1/T} d\tau e^{i(\omega+i\delta)\tau} \langle T_\tau Q_x(q,\tau)j_x(q,\tau)\rangle,$$ where $Q_i =  \sum_{\bk,n} \chi_n^\dagger(\bk)  E_n(\bk)\frac{\partial E_n(\bk)}{\partial k_i} \chi_n(\bk)$.
By taking $q \to 0$, $\omega \to 0$ and ignoring the inter-band contributions, we have the following results that are same as those obtained from the Boltzmann equation:
\bea
\label{conds}
&&
\alpha_{xx} =
-\frac{2e}{T}\sum_{\bk,n}
\tau(\bk)
(v_n^x)^2 E_n(\bk) \left( -\frac{\partial f(E_n(\bk))}{\partial E_n(\bk)}\right),
\\
&&
\sigma_{xx} =
2e^2\sum_{\bk,n}
\tau(\bk)
(v_n^x)^2 \left( -\frac{\partial f(E_n(\bk))}{\partial E_n(\bk)}\right),
\\
&&
\sigma_{xy} =
\frac{2e^3B}{\hbar c}\sum_{\bk,n}
\tau^2(\bk)
[ v_{n}^{x} v_{n}^{y} v_{n}^{xy} - (v_{n}^{x})^2 v_{n}^{yy}]\left( -\frac{\partial f(E_n(\bk))}{\partial E_n(\bk)}\right),
\eea
where $v_{n}^x = \frac{\partial E_n(\bk)}{\partial k_x}$ and
$v_{n}^{xy} = \frac{\partial^2 E_n(\bk)}{\partial k_x \partial k_y}$.

For a given doping concentration $p$, the temperature dependence of the CDW order parameter $W(p,T)$ is assumed to scale with a mean-field exponent below the critical temperature $T_{CDW}(p)$ such that $W(p,T) =  W(p) \left\vert 1 - T/T_{CDW}(p) \right\vert^{1/2}$ for $T<T_{CDW}(p)$, whereas $W(p,T)=0$ for $T>T_{CDW}(p)$.
Here, we used the phenomenological form for $T_{CDW}(p) = 142.6 (\text{K}) \times (p-p_{CDW})^{0.3}$ (where $p_{CDW} = 0.085$), which, although arbitrary, gives a best fit to the experiment~\cite{Laliberte:2011}.
Further, we assume the doping dependence of the CDW order parameter $W(p)$ below the critical temperature $T_{CDW}(p)$ such that it disappears below (above) a lower (upper) critical doping, producing a CDW dome consistent with the experiments. On the lower doping side we take $W(p) = W \left\vert (p-p_{CDW})/(p_\text{max}-p_{CDW}) \right\vert^{1/2}$ for $p> p_{CDW}$, and $W(p) = 0$ for $p<p_{CDW}$, and similarly for the higher doping side. Here, we set $p_\text{max} = 0.125$, and $W=0.069$ eV.
Although we use these values in our numerical calculations, we emphasize that our results are robust to reasonable variations of these parameters in the underdoped regime.
We notice that as CDW disappears, the Fermi surface becomes the contour of the dispersion $\varepsilon_\bk - \mu = 0$, which is closed around the corners of the Brillouin zone (the $M$ points), giving rise to the usual hole-like Fermi pockets.

With the conductivities given in Eqs.~(7,8,9), and the $T$ and $p$ dependences of CDW in hand, we calculate the Seebeck and Hall coefficients ($S = \alpha_{xx}/\sigma_{xx}$ and $R_H = \sigma_{xy}/\sigma_{xx}^2$, respectively) in the underdoped regime.
Fig.~\ref{fig-temp_dep} shows the Seebeck coefficient normalized by temperature, $S/T$, and the Hall coefficient $R_H$ as a function of $T$ for various doping concentrations $p$. As the temperature lowers, in the presence of CDW, the signs change from positive to negative in both Seebeck and Hall coefficients, while those for $p=0.08$ ($p<p_{CDW}$) remain positive. This indicates that the emergence of the electron-like pockets due to presence of the CDW gives rise to the negative Seebeck and Hall coefficients at low temperatures.

\begin{figure}[t]
\includegraphics[width = \linewidth]{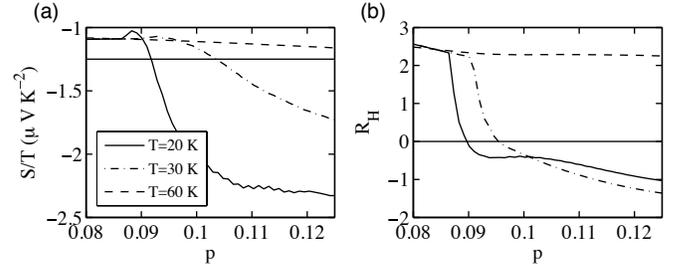}
\caption{
Doping dependences of Seebeck and Hall coefficients in the bi-axial CDW state with $Q_1=(2\pi/3,0)$ and $Q_2=(0,2\pi/3)$.
(a) With increasing doping concentration $p$, Seebeck coefficients $S/T$ for $T=20$ K and $30$ K change the sign from positive to negative, while $S/T$ at $T=60$ K ($W=0$) remains positive throughout.
(b) Hall coefficients $R_H$ vs. $p$ show the similar behavior as the doping dependence of Seebeck coefficient $S/T$.}
\label{fig-dop_dep}
\end{figure}

In Fig.~\ref{fig-dop_dep} we show the doping dependence of the Seebeck and Hall coefficients near the lower critical doping in the CDW phase (analogous results, not shown here, are found also near the higher critical doping).
Fig.~\ref{fig-dop_dep}a (\ref{fig-dop_dep}b) shows $S/T$ ($R_H$) as a function of doping concentration $p$ for various temperatures $T$. For a given temperature $T < T_{CDW}$, the positive Seebeck and Hall coefficients for $p<p_{CDW}$ ($W(p)=0$ for $p$ below the lower critical doping) becomes negative on crossing the critical doping concentration $p_{CDW}$, where the electron-like pockets emerge due to the CDW. When temperature is above the CDW critical temperature $T_{CDW}$, the Seebeck and Hall coefficients remain positive due to the hole-like Fermi surface around the $M$ points in the first Brillouin zone. Such temperature and doping dependences in the Seebeck and Hall coefficients (specifically, the sign changes at low temperatures in the presence of a bi-axial CDW) are qualitatively consistent with the experiments.

\section{Uni-axial CDW with $Q=(2\pi/3,0)$ or $(0,2\pi/3)$\label{seciv}}
\begin{figure}[b]
\includegraphics[width = 1\linewidth]{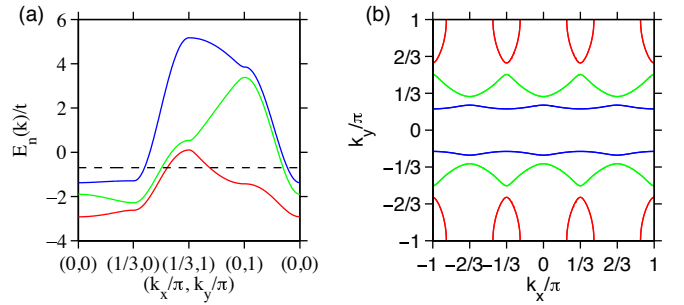}
\caption{
(color online)
(a) The CDW band structure from the eigenvalues $E_n(\bk)$ of the $3\times 3$ matrix in Eq.~(\ref{ham_stripe}). The chemical potential, corresponding to hole doping $p=0.125$, is displayed by a dashed line ($\mu = -1.0027 t$ for $W = 0.2267 t$ at $T=0$).
(b) The Fermi surface in the first Brillouin zone is depicted.
In this case, the reduced Brillouin zone (RBZ) boundary is determined by the CDW momentum $Q=(2\pi/3,0)$. Note that the Fermi surface in the RBZ consists of a hole pocket centered at $(\pi/3,\pi)$ depicted in red.
}
\label{fermi_surface_stripe}
\end{figure}
In this section, we show that a uni-axial CDW state with $Q=(2\pi/3,0)$ or $(0,2\pi/3)$ cannot produce such temperature or doping dependent crossover in sign of the Seebeck and Hall coefficients. This result is a consequence of the fact that reconstructed Fermi surface of uni-axial CDW with $Q=(2\pi/3,0)$ or $(0,2\pi/3)$ consists of only hole pockets and thus the Hall and Seebeck coefficients remain positive even at low temperatures.

To calculate the Fermi surface and the normal-state quasiparticle Hall and Seebeck coefficients in a uni-axial CDW with $\bqq=(2\pi/3,0)$ or $(0,2\pi/3)$, we use a 2D square lattice tight-binding dispersion, Eq.~(\ref{dispersion}).
\begin{figure}[t]
\includegraphics[width = \linewidth]{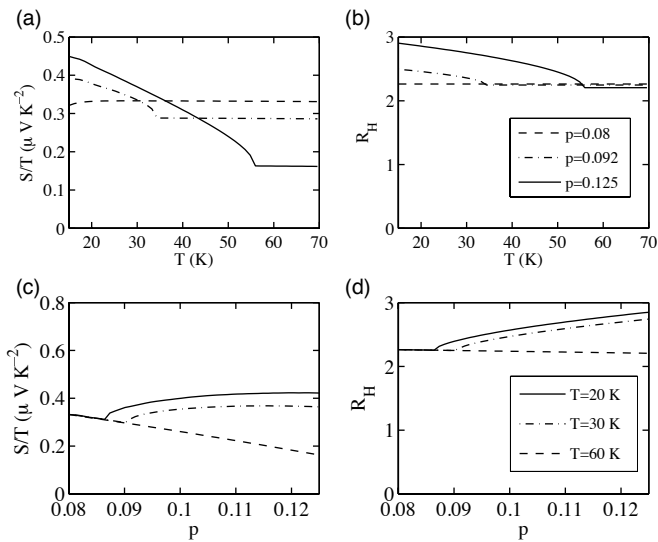}
\caption{
Temperature and doping dependences of the Seebeck and Hall coefficients in the uni-directional $Q = (2\pi/3,0)$ CDW state.
(a) and (b) the temperature dependence for $p=0.08$, 0.092, and 0.125.
(c) and (d) the doping dependence for $T=20$ K, 30 K, and 60 K.
}
\label{fig:coeff}
\end{figure}
Then, a uni-axial CDW with $\bqq=(2\pi/3,0)$ or $(0,2\pi/3)$ is given by a periodic density modulation  $\delta \rho (\mathbf{r}) = W\cos (\mathbf{Q}\cdot \mathbf{r})$, where $\mathbf{r} = (x,y)$.
In the reciprocal momentum space the Fourier transform $W_\bq$ can be expressed as
$
W_\bq = W \delta_{\bq,\mathbf{Q}}.
\label{h_stripe}
$
The mean-field Hamiltonian can be written as
\begin{eqnarray}
H_{MF}
&=&
\sum_{\bk,\sigma} \left( \varepsilon_\bk - \mu \right) c_{\bk\sigma}^\dagger c_{\bk\sigma}+
W \sum_{\bk,\sigma}  c_{\bk+\bqq,\sigma}^\dagger c_{\bk\sigma} + H.c.
\nonumber \\
&=&\frac{1}{2}
\sum_{\bk \in RBZ}
\psi_\bk^\dagger
\bmatrix
\xi_{\bk-\mathbf{Q}} & W & W \\
W & \xi_{\bk} & W \\
W & W & \xi_{\bk+\mathbf{Q}}
\ematrix
\psi_{\bk},
\label{ham_stripe}
\end{eqnarray}
where $RBZ$ is $|k_x|\le \frac{\pi}{3}, |k_y|\le \pi$ and  $\psi_{\bk}^\dagger = [ c_{\bk-\mathbf{Q}} , c_{\bk}, c_{\bk+\mathbf{Q}} ]$.
Fig.~\ref{fermi_surface_stripe} illustrates (a) band structure and (b) Fermi surface in the 1st Brillouin zone at $p=0.125$ and $T=0$.

We calculate the Hall and Seebeck coefficients in the presence of a uni-axial CDW with $\bqq=(2\pi/3,0)$ as functions of temperature $T$ and doping $p$ in the underdoped regime (Fig.~\ref{fig:coeff}) with the same parameters as in the previous sections:$p_{CDW}=0.085$, $p_{\text{max}}=0.125$, $W=0.069$ eV, and $T_{CDW}(p) = 142.6 (\text{K}) \times (p-p_{CDW})^{0.3}$.
Due to the unidirectional feature of stripe CDW, the Seebeck $S$ and Hall coefficients $R_H$ depend upon the directions of the applied thermal gradient and transverse electric field generated by the magnetic field: When parallel to $Q=(2\pi/3,0)$, or perpendicular to $Q$.
Contrary to the bi-axial with $Q_1$ and $Q_2$ CDW state, in both cases, the sign change with onset of CDW order parameter in  the coefficients $S$ and $R_H$ does not occur. Positive $S$ and $R_H$ at high temperatures or $p<p_{CDW}$ ($W=0$) remain positive even in the presence of this uni-axial CDW order parameter. It can be explained by the presence of the hole pockets in the Fermi surface in the uni-axial CDW state with $\bqq=(2\pi/3,0)$ or $(0,2\pi/3)$.

\section{Conclusion\label{secv}}
Motivated by strong experimental evidence of an incipient charge density wave instability in the underdoped regime of high-$T_c$ cuprates, we calculate the Fermi surface topology and the resulting Hall and Seebeck coefficients for a mean field $Q_1=(2\pi/3,0)$ and $Q_2=(0,2\pi/3)$ bi-axial CDW state. We show that, in the appropriate doping ranges in which the low temperature state (in the absence of superconductivity) is a bi-axial CDW, the Fermi surface consists of electron and hole pockets, resulting in the Hall and Seebeck coefficients becoming negative at low temperatures. Our calculated temperature and doping dependences of the transport coefficients, specifically a change of sign at low temperatures in a restricted range of hole-doping, are qualitatively consistent with experiments.
In addition to explaining the temperature dependent crossover of the magneto-transport coefficients in the underdoped regime of cuprates in terms of the recently observed CDW state, our results also show that the uni-axial CDW state is inconsistent with the phenomenology of transport coefficients in the cuprates at high magnetic fields.

\begin{acknowledgments}
{\it This work is supported by
NSF-PHY (1104527), and AFOSR (FA9550-13-1-0045).}
\end{acknowledgments}




\begin{thebibliography}{99}

\bibitem{Norman:2005} M. R. Norman, D. Pines, C. Kallin, Adv. Phys. \textbf{54}, 715 (2005).

\bibitem{Varma:1997} C. M. Varma, Phys. Rev. B  {\bf 55}, 14554 (1997); {\em ibid} {\bf 73}, 155113 (2006).

\bibitem{Chakravarty:2001} S. Chakravarty, R. B. Laughlin, D. K. Morr, C. Nayak, Phys. Rev. B \textbf{63}, 094503 (2001).

\bibitem{Kivelson:2003} S. A. Kivelson, I. P. Bindloss, E. Fradkin, V. Oganesyan, J. M. Tranquada, A. Kapitulnik, and C. Howald, Rev. Mod. Phys. {\bf 75}, 1201 (2003).

\bibitem{Hosur:2012}Pavan Hosur, A. Kapitulnik, S.A. Kivelson, J. Orenstein, and S. Raghu, Phys. Rev. B \textbf{87}, 115116 (2013).

\bibitem{Orenstein:2012}J. Orenstein and  Joel E. Moore, Phys. Rev. B \textbf{87}, 165110 (2013).

\bibitem{Chakravarty:2013}S. Chakravarty, Phys. Rev. B \textbf{89}, 087101 (2014).

\bibitem{Wu:2011} T. Wu, H. Mayaffre, S. Kr\"{a}mer, M. Horvati\'{c}, C. Berthier, W. N. Hardy, R. Liang, D. A. Bonn, and M.-H. Julien, Nature \textbf{477}, 191--194 (2011).

\bibitem{Wu:2013} T. Wu, H. Mayaffre, S. Kr\"{a}mer, M. Horvati\'{c}, C. Berthier, P. L. Kuhns, A. P. Reyes, R. Liang, W. N. Hardy, D. A. Bonn, and M.-H. Julien, Nature Communications \textbf{4}, 2113 (2013).

\bibitem{Doiron-Leyraud:2007} N. Doiron-Leyraud, C. Proust, D. LeBoeuf, J. Levallois, J.-B. Bonnemaison, R. Liang, D. A. Bonn, W. N. Hardy, L. Taillefer,
Nature \textbf{447}, 565568 (2007).

\bibitem{Sebastian:2008} S. E. Sebastian, N. Harrison, E. Palm, T. P. Murphy, C. H. Mielke, R. Liang, D. A. Bonn, W. N. Hardy, G. G. Lonzarich, Nature \textbf{454}, 200 (2008).
\bibitem{LeBoeuf:2007} D. LeBoeuf, N. Doiron-Leyraud, J. Levallois, R. Daou, J.-B. Bonnemaison, N. E. Hussey, L. Balicas, B. J. Ramshaw, R. Liang, D. A. Bonn, W. N. Hardy, S. Adachi, C. Proust, L. Taillefer, Nature \textbf{450}, 533 (2007).

    \bibitem{Chang:2010} J. Chang,  R. Daou, C. Proust, D. LeBoeuf, N. Doiron-Leyraud, F. Lalibert\'{e}, B. Pingault, B. J. Ramshaw, R. Liang, D. A. Bonn, W. N. Hardy, H. Takagi, A. B. Antunes, I. Sheikin, K. Behnia, and L. Taillefer, Phys. Rev. Lett. \textbf{104}, 057005 (2010).

   \bibitem{Laliberte:2011} F. Lalibert\'{e}, J. Chang, N. Doiron-Leyraud, E. Hassinger, R. Daou, M. Rondeau, B. J. Ramshaw, R. Liang, D. A. Bonn, W. N. Hardy, S. Pyon, T. Takayama, H. Takagi, I. Sheikin, L. Malone, C. Proust, K. Behnia, and L. Taillefer, Nature Communications \textbf{2}, 432 (2011).



\bibitem{Chakravarty-Kee} S. Chakravarty, H.-Y. Kee,  Proc. Natl. Acad. Sci. USA \textbf{105}, 8835 (2008).

\bibitem{Millis-Norman} A. J. Millis, M. R. Norman, Phys. Rev. B \textbf{76}, 220503(R) (2007).



\bibitem{Chang:2012} J. Chang, E. Blackburn, A. T. Holmes, N. B. Christensen, J. Larsen, J. Mesot,
R. Liang, D. A. Bonn, W. N. Hardy, A.Watenphul, M. v. Zimmermann, E. M. Forgan,
S. M. Hayden, Nature Phys. \textbf{8}, 871 (2012).

\bibitem{Ghiringhelli:2012} G. Ghiringhelli, M. Le Tacon, M. Minola, S. Blanco-Canosa, C. Mazzoli,
N. B. Brookes, G. M. De Luca, A. Frano, D. G. Hawthorn, F. He, T. Loew,
M. Moretti Sala, D. C. Peets, M. Salluzzo, E. Schierle, R. Sutarto, G. A. Sawatzky,
E. Weschke, B. Keimer, L. Braicovich, Science \textbf{337}, 821 (2012).

\bibitem{Kivelson} E. Fradkin, S. A. Kivelson, Nature Physics \textbf{8}, 865 (2012).


\bibitem{Tranquada}J. M. Tranquada, Science \textbf{337}, 811 (2012).

\bibitem{LeBoeuf:2013} D. LeBoeuf,	S. Kr\"{a}mer, W. N. Hardy, R. Liang, D. A. Bonn, and C. Proust, Nature Phys. \textbf{9}, 79 (2013).

\bibitem{Hinton:2013} J. P. Hinton, J. D. Koralek, Y. M. Lu, A. Vishwanath, J. Orenstein, D. A. Bonn, W. N. Hardy, and R. Liang, Phys. Rev. B \textbf{88}, 060508(R) (2013).

\bibitem{sebastian:2014} S. E. Sebastian, N. Harrison, F. F. Balakirev, M. M. Altarawneh, P. A. Goddard,	R. Liang, D. A. Bonn, W. N. Hardy, and G. G. Lonzarich, Nature \textbf{511}, 61 (2014).

\bibitem{Fujita:2014} K. Fujita, M. H. Hamidian, S. D. Edkins, C. K. Kim, Y. Kohsaka, M. Azuma, M. Takano, H. Takagi, H. Eisaki, Shin-ichi Uchida, A. Allais, M. J. Lawler, E.-A. Kim, S. Sachdev, and J. C. S. Davis, PNAS \textbf{111}, 30 (2014).

\bibitem{Comin:2014} R. Comin, R. Sutarto, F. He, E. da Silva Neto, L. Chauviere, A. Frano, R. Liang, W. N. Hardy, D. Bonn, Y. Yoshida, H. Eisaki, J. E. Hoffman, B. Keimer, G. A. Sawatzky, A. Damascelli, arXiv:1402.5415.

\bibitem{sebastian} S. E. Sebastian, N. Harrison, and G. G. Lonzarich, Rep. Prog. Phys. \textbf{75}, 102501 (2012).


%
\bibitem{hildebrand} G.~Hildebrand, T.~J.~Hagenaars, W.~Hanke, S.~Grabowski, and J.~ Schmalian, Phys. Rev. B \textbf{56}, 4317 (R) (1997).
%
\bibitem{kontani} Hiroshi Kontani, Rep. Prog. Phys. \textbf{71}, 026501 (2008)
%
\bibitem{storey} J.~G.~Storey, J.~L.~Tallon, and G.~V.~M.~Williams, EPL \textbf{102}, 37006 (2013).
%

%
\end{thebibliography}
\end{document}